\pdfoutput=0
\documentclass[tightenlines,aps,floatfix,preprint,
nofootinbib]{revtex4}

\usepackage{float,ifpdf}
\usepackage{amsmath,amsfonts,axodraw,graphicx,bm}
\usepackage{times}

\newcommand{\ep}{\epsilon}
\newcommand{\be}{\begin{equation}}
\newcommand{\ee}{\end{equation}}
\newcommand{\ba}{\begin{eqnarray}}
\newcommand{\ea}{\end{eqnarray}}

\begin{document}

%\begin{titlepage}

\begin{flushright}
\vbox{
\begin{tabular}{l}
UH-511-1101-07
\end{tabular}
}
\end{flushright}

\title{
QCD corrections to tri-boson production
}
\author{Achilleas Lazopoulos}
\author{Kirill Melnikov
        \thanks{e-mail: kirill@phys.hawaii.edu}}
\affiliation{Department of Physics and Astronomy,
          University of Hawaii,\\ 2505 Correa Rd. Honolulu, HI 96822}  
\author{Frank Petriello\thanks{frankjp@phys.wisc.edu}}
\affiliation{
Department of Physics, University of Wisconsin, Madison, WI  53706
\medskip
} 

\begin{abstract}
\medskip
We present a computation of the next-to-leading order QCD corrections to the production 
of three $Z$ bosons at the LHC.  We calculate these corrections using 
a completely numerical method that combines sector decomposition to extract infrared singularities
with contour deformation of the Feynman parameter integrals to avoid internal loop thresholds. The NLO QCD corrections 
to $pp \to ZZZ$ are approximately 50\%, and are badly underestimated by the leading order scale dependence.  However, 
the kinematic dependence of the corrections is minimal in phase space regions accessible at leading order.

\end{abstract}

\maketitle

%\thispagestyle{empty}
%\end{titlepage}

\section{Introduction}
\label{sect1}

The search for and interpretation of new physics at the LHC will 
require a precise understanding 
of the Standard Model.  Without accurate QCD predictions and reliable error estimates 
for important observables, mistakes in interpreting experimental results may occur.  Notable 
recent examples where poor theoretical understanding has hindered the analysis of an 
experimental result are the deviation of the Brookhaven muon $g-2$ measurement from the 
Standard Model prediction, the excess of bottom quark production in Run I of the Tevatron, 
and the discrepancy in the weak mixing angle obtained by NuTeV.  At the LHC, all analyses require 
perturbative calculations to at least next-to-leading order (NLO) in $\alpha_s$, the QCD coupling 
constant, in order to make quantitative predictions that are free from debilitating 
theoretical uncertainties.  This need for higher order calculations at the LHC has been summarized 
in an experimental "NLO wishlist" of processes for which QCD corrections are 
desired~\cite{NLOwish}.

A cross section at higher orders in perturbation theory consists of two primary components: 
virtual corrections, in which additional loops are added to the Born-level matrix element, 
and real 
corrections, in which additional partons are radiated.  Each contribution
is separately infrared divergent.  They must be combined and summed over degenerate final 
states to obtain a finite result.  Initial-state collinear singularities must be absorbed 
into the definitions of the parton distribution functions.

Well-developed techniques exist for the calculation of real emission corrections at NLO.
However, the calculation  of 
virtual corrections to processes with many particles 
in the final state remains a challenge. For this reason, 
our primary focus in this paper will be to develop an approach to computing virtual corrections 
to $2 \to 3$ scattering processes. We begin by discussing methods currently used to perform such 
calculations.  Well-developed techniques exist for calculating one-loop
virtual corrections to $2 \to 2$ 
and simpler processes.  The matrix elements are obtained via a standard Feynman diagram 
calculation.  The tensor integrals are reduced to a basis of scalar integrals via a 
reduction algorithm such as Passarino-Veltman~\cite{Passarino:1978jh}.  The basis integrals are 
then computed analytically using a Feynman parameter representation, with care being 
taken to extract all infrared singularities that occur in the parametric integration.  These 
integrals are typically performed with Euclidean kinematics; after the analytic expression is 
derived, the resulting logarithms and polylogarithms are analytically continued to the 
Minkowski region.

Several problems arise when this approach is extended to $2 \to 3$ and more complicated scattering 
processes.  The increase in algebraic complexity alone makes the procedure difficult.  The singularity 
extraction and analytic continuation to the physical region are typically done on a case-by-case 
basis for each scattering process; the large number of processes 
for which NLO QCD computations are desired implies that studying 
each separately will be an enormously time-consuming task.  Furthermore, the standard reduction algorithms introduce inverse 
Gram determinants multiplying the basis integrals; these coefficients vanish when the final-state momenta become linearly dependent,
 and can become arbitrarily small nearby.  Although these spurious singularities cancel when all basis integrals are combined, it is 
difficult to establish this analytically, and they usually cause serious numerical complications.  
Careful studies of the boundaries and extrapolations of numerical results from safe phase space 
regions are typically required to obtain stable answers~\cite{NLOexamples}.

Because of these complications, the computation of the NLO QCD corrections to $2 \to n, n\geq 3$ scattering 
processes is a difficult and intricate task that typically requires one year or more of effort for 
every interaction considered.  These difficulties and the importance of these calculations to the 
LHC physics program 
have stimulated significant effort to develop new approaches for perturbative 
QCD computations.  These include new analytic~\cite{NLOan} and semi-numerical~\cite{NLOnum} methods 
for evaluating loop integrals.  
Phenomenological results obtained with these methods
include the NLO QCD corrections to $H+2~{\rm jets}$~\cite{Campbell:2006xx} 
and $t\bar{t} + {\rm jet}$~\cite{Dittmaier:2007wz} 
at the LHC, and also the matrix elements used for $W,Z+{\rm jets}$~\cite{zvi}.

Ideally, an algorithm for NLO QCD calculations would be highly automated and would handle a large 
number of processes without the need to consider special cases.  This would allow a large swath of 
desired corrections to be computed by a single program running in parallel on many machines.  Any such 
approach must confront three main issues: 
spurious phase space singularities that appear during the
reduction of tensor integrals, the extraction of soft and collinear singularities, 
and the presence of internal thresholds where analytic 
continuation is required.  An 
approach that addresses the first two issues exists, called sector decomposition~\cite{Hepp:1966eg,Roth:1996pd,Binoth:2000ps}.  
It permits a completely automated, numerical extraction of infrared singularities from loop integrals.  The 
application of this approach to an integral
results in 
a Laurent series in $\ep$, the parameter of dimensional regularization, with coefficients that can be numerically integrated over 
Feynman parameters.  Since the infrared singularity structure of a loop diagram in parametric 
space is completely determined by its denominator, no Passarino-Veltman
reduction of tensor integrals is needed.  Consequently, inverse Gram determinants never appear.  The 
basic scalar integral which characterizes a diagram is identified and sector decomposed.  The tensors become 
polynomials in the Feynman parameters after integration over the loop momenta; they can be treated numerically.  

The remaining issue is the internal threshold structure present in loop diagrams.  Thresholds occur when the internal 
propagators go on-shell, and a unitarity cut of the diagram leads to a physical scattering process.  In Feynman 
parameter space, the denominator vanishes at these points, and is regulated only by the $-i0$ prescription for loop integrals.  
For $N$-point functions, this leads to denominators with the behavior $1/(-i0)^{N-2}$ at threshold locations.  This is completely 
unsuitable for numerical implementation.  An approach for handling thresholds in loop diagrams numerically was 
developed in~\cite{Soper,Soper2}.  It entails a 
contour deformation of the Feynman parametric integrals off the real axis and into the complex plane to 
avoid internal thresholds.  The integrals are then computed numerically.  The choice of 
the deformation for a given diagram is easily automated.

It appears to us that the combination of sector decomposition and contour deformation provides a framework 
in which numerical
calculations of NLO virtual corrections can be fully automated.
A similar attitude was espoused 
in~\cite{Binoth:2005ff}.  Our goal in this 
paper is to test this idea on a realistic $2 \to 3$ scattering process at the LHC.  We study the NLO QCD 
corrections to $pp \to ZZZ+X$, which acts as a background to supersymmetric tri-lepton production and 
appears on the NLO wishlist in~\cite{NLOwish}.  We find that the combination of these procedures does indeed appear 
to be a convenient approach to NLO QCD computations.

This paper is organized as follows.  In Section~\ref{sect2} 
we define our notation and present $pp \to ZZZ$ at 
leading order in QCD.  In Section~\ref{sect3} 
we discuss our computation of the NLO QCD corrections.  In particular, 
we present the algorithm we use for the computation of the virtual corrections.  
In Section~\ref{sect4} we provide numerical results for 
$pp \to ZZZ+X$ at the LHC.  We conclude in Section~\ref{sect5}.

\section{Setup and leading order process}
\label{sect2}

We consider the production of three $Z$-bosons in proton-proton collisions,
\be
p(P_1) + p(P_2) \to Z(p_3)+Z(p_4)+Z(p_5) + X.
\ee
Within the framework of QCD factorization, 
the cross section for this process is 
\be
{\rm d} \sigma = \sum_{ij} \int_{0}^{1} dx_1 dx_2 f_i^{p_1}(x_1) f_j^{p_2}(x_2) 
{\rm d} \sigma_{ij \to 3Z+X}(x_1,x_2),
\label{baseform}
\ee
where the $f_{i}^{p_j}$ are parton distribution functions that 
describe the probability to find a parton $i$ with momentum $xP_j$ in 
the proton $p_j$. The partonic cross sections ${\rm d}\sigma_{ij}$ are 
computed perturbatively as an expansion in the strong coupling constant 
$\alpha_s$:
\be
{\rm d}\sigma_{ij} = {\rm d}\sigma^{(0)}_{ij} 
+ \left (\frac{\alpha_s}{\pi} \right) {\rm d}\sigma^{(1)}_{ij} 
+ {\cal O}(\alpha_s^2).
\label{asexp}
\ee

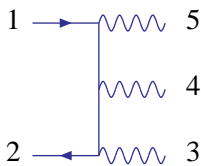
\begin{figure}[htb]
     \begin{center}
      \begin{picture}(50,50)(0,0)
        \SetColor{Blue}
        \ArrowLine(0,50)(25,50)
        \Photon(25,50)(50,50){3}{4}
       \ArrowLine(25,0)(0,0)
       \Photon(25,0)(50,0){3}{4}
       \Line(25,0)(25,50)
        \Photon(25,25)(50,25){3}{4}
        \put(-10,-2){2}
        \put(-10,48){1}
        \put(58,-2){3}
        \put(58,48){5}
        \put(58,23){4}
      \end{picture}
     \end{center}
\caption{\label{born} 
          Representative Born-level diagram for $q\bar{q} \to ZZZ$.}
\end{figure}

At leading order in this expansion, only the partonic channel $q(p_1) + \bar{q}(p_2) \to Z(p_3)+Z(p_4)+Z(p_5)$ 
contributes.  A representative diagram for this process is shown in Fig.~(\ref{born}).  There are six such diagrams, which can 
be obtained via permutation of the final-state bosons.  We neglect diagrams containing the exchange of a Higgs boson.
For Higgs boson masses below $2M_Z$, the contributions from these diagrams are small.  At next-to-leading order, both 
virtual corrections and additional radiative processes occur; we discuss the calculation of these components in later sections.  
After combining the virtual and real corrections, the partonic cross sections in Eq.~(\ref{asexp}) contain collinear singularities 
arising from initial-state radiation.  These are absorbed into the definitions of the parton distribution functions, as discussed 
in a later section.

The expression for the leading order cross section is
\be
{\rm d}\sigma^{(0)}_{q\bar{q}} = \frac{1}{4}\frac{1}{9}\frac{1}{6}\frac{1}{2\hat{s}} |{\cal M}^{(0)}|^2 {\rm d}\Omega_{3Z},
\ee
where the factors $\frac{1}{4}$, $\frac{1}{9}$, and $\frac{1}{6}$ are from spin-averaging, color-averaging, and identical particles, 
$\hat{s} =x_1 x_2 s$ is the partonic center-of-momentum energy squared, 
$s = 2P_1\cdot P_2$ is the total energy squared of the proton-proton collision
and $\Omega_{3Z}$ denotes the final-state phase space.  The matrix 
elements $|{\cal M}^{(0)}|^2$ have the expansion
\be 
|{\cal M}^{(0)}|^2 = |{\cal M}_{0}^{(0)}|^2+\ep |{\cal M}_{1}^{(0)}|^2,
\label{LOepexp}
\ee
where $d=4-2\ep$ is the space-time dimensionality in dimensional regularization.  The matrix elements are simple to calculate
using standard Feynman diagram techniques.  We use a combination of the programs QGRAF~\cite{Nogueira:1991ex}, FORM~\cite{Vermaseren:2000nd}, 
and MAPLE to obtain them.  The required electroweak vertex is
\begin{eqnarray}
Z\bar{q}{q}: & & i \sqrt{\frac{8M_W^2 G_F}{\sqrt{2}c_w}} \left(g_v+g_a \gamma_5\right), \nonumber \\ 
& & g_v = \frac{T_3^q}{2}-Q_q s_w^2,\,\,\, g_a = -\frac{T_3^q}{2}.
\end{eqnarray}
Here, $s_w^2 = 1-M_W^2/M_Z^2$ is the sine squared of the electroweak mixing angle, $T_3^q$ is the weak isospin of the 
quark $q$, $Q_q$ is the electric charge of the quark $q$ in units of the proton charge, and $G_F$ is the Fermi constant.  
Decomposing the matrix elements using the expansion in Eq.~(\ref{LOepexp}), 
the leading order cross section takes the form
\be
{\rm d}\sigma^{(0)} ={\rm d}\sigma_{0}^{(0)} + \ep {\rm d}\sigma_{1}^{(0)}.
\label{LOepexp2}
\ee
The ${\cal O}(\ep)$ 
term in Eq.~(\ref{LOepexp2}) is needed in the computation of the collinear counterterms.

\section{Next-to-leading order corrections}
\label{sect3}

The ${\cal O}(\alpha_s)$ NLO QCD corrections consist of the following components:
\begin{enumerate}
 \item the radiative processes $q\bar{q} \to ZZZg$, $qg \to ZZZq$, and $\bar{q}g \to ZZZ\bar{q}$;
 \item the collinear counterterms which absorb the initial-state collinear singularities of these 
   radiative processes into the parton distribution functions;
 \item the one-loop virtual contributions to the leading order partonic process $q\bar{q} \to ZZZ$.
\end{enumerate}
\noindent We discuss the calculation of each component in the following sections, and 
describe in detail the method used to compute the virtual corrections.

\subsection{Real radiation}

We begin with a discussion of the real radiation processes.  
We present in detail the process $q\bar{q} \to ZZZg$, and 
then note the modifications required when $qg \to ZZZq$ and $\bar{q}g \to ZZZ\bar{q}$ are considered.  

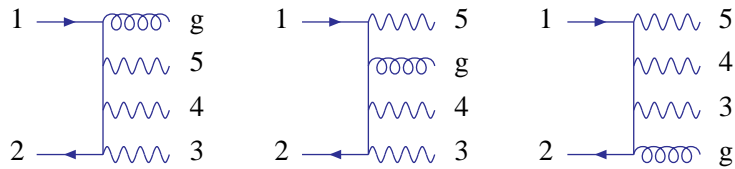
\begin{figure}[htb]
     \begin{center}
      \begin{picture}(230,50)(0,0)
        \SetColor{Blue}
        \ArrowLine(0,50)(25,50)
        \ArrowLine(25,0)(0,0)
        \Line(25,0)(25,50)
        \Gluon(25,50)(50,50){3}{4}
        \Photon(25,0)(50,0){3}{4}
        \Photon(25,16.7)(50,16.7){3}{4}
        \Photon(25,33.4)(50,33.4){3}{4}
        \put(-10,-2){2}
        \put(-10,48){1}
        \put(58,-2){3}
        \put(58,48){g}
        \put(58,31.4){5}
        \put(58,14.7){4}

        \ArrowLine(100,50)(125,50)
        \ArrowLine(125,0)(100,0)
        \Line(125,0)(125,50)
        \Photon(125,50)(150,50){3}{4}
        \Photon(125,0)(150,0){3}{4}
        \Photon(125,16.7)(150,16.7){3}{4}
        \Gluon(125,33.4)(150,33.4){3}{4}
        \put(90,-2){2}
        \put(90,48){1}
        \put(158,-2){3}
        \put(158,48){5}
        \put(158,31.4){g}
        \put(158,14.7){4}

	\ArrowLine(200,50)(225,50)
        \ArrowLine(225,0)(200,0)
        \Line(225,0)(225,50)
        \Photon(225,50)(250,50){3}{4}
        \Gluon(225,0)(250,0){3}{4}
        \Photon(225,16.7)(250,16.7){3}{4}
        \Photon(225,33.4)(250,33.4){3}{4}
        \put(190,-2){2}
        \put(190,48){1}
        \put(258,-2){g}
        \put(258,48){5}
        \put(258,31.4){4}
        \put(258,14.7){3}

      \end{picture}
     \end{center}
\caption{\label{NLORdiags} 
          Representative diagrams contributing to $q\bar{q} \to ZZZg$.}
\end{figure}

Twenty-four Feynman 
diagrams contribute to $q(p_1)+\bar{q}(p_2) \to Z(p_3)+Z(p_4)+Z(p_5)+g(p_g)$.  A few representative samples are shown in Fig.~(\ref{NLORdiags}).  
Collinear and soft singularities occur when the gluon is emitted from an initial fermion line, indicating that the Laurent series for this process 
begins at $1/\ep^2$.  The expression for the cross section is
\be
{\rm d}\sigma^{(1)}_{q\bar{q} \to 3Z+g} = \frac{1}{4} \frac{1}{9} \frac{1}{6} \frac{1}{2\hat{s}} |{\cal M}_{3Z+g}^{(1)}|^2 {\rm d}\Omega_{3Z+g}.
\ee
The matrix elements are again simple to obtain using standard techniques.

To discuss the extraction of the infrared singularities, it is convenient to introduce an explicit parameterization of the final-state 
phase space.  In terms of the partonic momenta, the phase space takes the form
\begin{eqnarray}
\int {\rm d}\Omega_{3Z+g} &=& \int \frac{d^4 p_3}{(2\pi)^3} \frac{d^4 p_4}{(2\pi)^3}\frac{d^4 p_5}{(2\pi)^3} \frac{d^d p_g}{(2\pi)^{d-1}} 
 \,\delta(p_3^2-M_Z^2)\,\delta(p_4^2-M_Z^2)\,\delta(p_5^2-M_Z^2)\,\delta(p_g^2) \nonumber \\ &\times& (2\pi)^{d}\delta^{(d)}(p_1+p_2-p_3-p_4-p_5-p_g).
\end{eqnarray}
Since there are no singularities associated with the $3Z$ phase space, we 
evaluate it directly in four dimensions.  
We now partition the phase space 
using 
\begin{eqnarray}
&&\int \frac{d^4 p_3}{(2\pi)^3} \frac{d^4 p_4}{(2\pi)^3}\frac{d^4 p_5}{(2\pi)^3} \frac{d^d p_g}{(2\pi)^{d-1}} 
 \,\delta(p_3^2-M_Z^2)\,\delta(p_4^2-M_Z^2)\,\delta(p_5^2-M_Z^2)\,\delta(p_g^2) \nonumber \\ &\times& (2\pi)^{d}\delta^{(d)}(p_1+p_2-p_3-p_4-p_5-p_g) 
= \int \frac{ds_{345}}{2\pi} \int \frac{d^d p_{345}}{(2\pi)^{d-1}}  \frac{d^d p_g}{(2\pi)^{d-1}} \,\delta(p_{345}^2-s_{345})\,\delta(p_g^2) 
\nonumber \\ &\times& (2\pi)^{d}\delta^{(d)}(p_1+p_2-p_{345}-p_g) \int {\rm d}\Omega_{3Z}(p_{345}),
\end{eqnarray}
where $\Omega_{3Z}(p_{345})$ denotes the $3Z$ phase space with the sum of the three boson momenta giving $p_{345}$ rather than $p_1+p_2$.  We evaluate 
the gluon phase space in the partonic center-of-momentum frame by introducing the explicit four-momenta
\begin{equation}
p_1 = \frac{\sqrt{\hat{s}}}{2}\left(1,0,0,1\right),\,\,
p_2 = \frac{\sqrt{\hat{s}}}{2}\left(1,0,0,-1\right),\,\,
p_g = E_g\left(1,0,s_g,c_g\right).
\end{equation}
We use the $\delta$-functions to remove as many integrations as possible, and change variables in those remaining so that the boundaries are at 
0 and 1.  We arrive at the following expression for the phase space:
\be
\int {\rm d}\Omega_{3Z+g} =\frac{\pi^{1-\ep}}{2(2\pi)^{d-1}\Gamma(1-\ep)} \int_{0}^{1} d\lambda_1 d\lambda_5 \left[\lambda_1(1-\lambda_1)\right]^{-\ep} 
 \lambda_{5}^{1-2\ep}\left(1-9z^2\right)^{2-2\ep} \int {\rm d}\Omega_{3Z}(p_{345}).
\label{NLORPS}
\ee
Here, $z^2 = M_Z^2/\hat{s}$, and the expressions for the invariant masses in terms of the hypercube variables $\lambda_1$ and $\lambda_5$ are
\begin{eqnarray}
s_{345} &=& (p_3+p_4+p_5)^2 = (1-9z^2)(1-\lambda_5)+9z^2, \nonumber \\ 
s_{1g} &=& (p_1-p_g)^2 = -\lambda_5(1-\lambda_1)(1-9z^2), \nonumber \\ 
s_{2g} &=& (p_2-p_g)^2 = -\lambda_5\lambda_1(1-9z^2).
\label{NLORinvs}
\end{eqnarray}
In writing these expressions we have set the overall energy scale $\hat{s}=1$; it can be restored at the end using dimensional analysis.

The singular terms in the matrix elements come from the following three sources.
\begin{enumerate}
 \item Interferences between diagrams where the gluon is emitted from 
the quark line with momentum $p_1$.  When the denominator of the off-shell quark propagator
  $s_{1g}$ in Eq.~(\ref{NLORinvs}) is combined with the phase space in Eq.~(\ref{NLORPS}), the singular structure $\lambda_{5}^{-1-2\ep} 
  (1-\lambda_1)^{-1-\ep}$ is obtained.
 \item Interferences between diagrams where the gluon is emitted from the anti-quark line with momentum
$p_2$: these 
  lead to the singular structure $\lambda_{5}^{-1-2\ep}\lambda_{1}^{-1-\ep}$.  
 \item Interferences containing the denominator $1/s_{1g}/s_{2g}$: if care is taken to sum over only physical gluon polarizations 
   in the final state, these contain only the soft singularity $\lambda_{5}^{-1-2\ep}$.
\end{enumerate}
To extract the singularities as a Laurent series in $\ep$, we use the following standard plus distribution expansion:
\be
\lambda^{-1+\ep} = \frac{1}{\ep}\delta(\lambda) +\sum_{n=0} \frac{\ep^n}{n!} \left[\frac{{\rm ln}^n \lambda}{\lambda}\right]_{+},
\label{plusexp}
\ee
where the plus distributions are defined via
\be
\int_{0}^{1} d\lambda \left[\frac{f(\lambda)}{\lambda}\right]_{+} g(\lambda) = \int_{0}^{1} d\lambda \frac{f(\lambda)}{\lambda} 
 \left(g(\lambda)-g(0)\right).
\ee
After using these expansions, the cross section for $q\bar{q} \to ZZZg$ takes the form
\be
{\rm d}\sigma^{(1)}_{q\bar{q} \to 3Z+g} = \frac{{\cal A}_{2}}{\ep^2}+\frac{{\cal A}_{1}}{\ep}+{\cal A}_{0}.  
\label{NLORepexp}
\ee
The ${\cal A}_i$ are integrable, $\ep$-independent quantities that contain the complete kinematic information of the final state.  
The $1/\ep^2$ singularities, 
where $\lambda_5 \to 0$ and $\lambda_1 \to 0$ or $1$, cancel against the virtual contributions to $q\bar{q} \to ZZZ$.  The $1/\ep$ terms where 
$\lambda_1 \to 0$ or $1$ are removed by the collinear counterterms discussed in the next section.  The $1/\ep$ singularities 
where $\lambda_5 \to 0$ cancel against a combination of the virtual corrections and collinear counterterms.

The matrix elements for the remaining real radiation processes $qg \to ZZZq$ and $\bar{q}g \to ZZZ\bar{q}$ are identical.  They each consist 
of twenty-four diagrams.  The cross section is
\be
{\rm d}\sigma^{(1)}_{qg \to 3Z+q} = \frac{1}{4(1-\ep)} \frac{1}{24} \frac{1}{6} \frac{1}{2\hat{s}} |{\cal M}_{3Z+q}^{(1)}|^2 {\rm d}\Omega_{3Z+q},
\ee
where we have used the fact that
the gluon has $2(1-\ep)$ physical polarizations in $d=4-2\ep$ dimensions.  We extract singularities using the same phase 
space parameterization and expansion 
in plus distributions discussed above.  For this process, only collinear singularities where $\lambda_1 \to 0$ or $1$ occur.  These 
are removed by the collinear counterterms described in the next section.

\subsection{Collinear counterterms}

The radiative processes discussed in the previous section contain collinear singularities that must be absorbed into the parton distribution 
functions.  To do so, we begin by expressing the bare distribution functions and cross sections in Eq.~(\ref{baseform}) in terms of the renormalized ones,
\be
{\rm d} \sigma = \sum_{ij} \int_{0}^{1} dx_1 dx_2 \hat{f}_i^{p_1}(x_1) \hat{f}_j^{p_2}(x_2) 
{\rm d} \hat{\sigma}_{ij}(x_1,x_2).
\label{newform}
\ee
The renormalized parton distribution functions $\hat{f}_i$ are related to the bare ones $f_i$ via
\be
\hat{f}_i = \Gamma_{ij} \otimes f_j.
\ee
We have introduced the convolution integral
\be
(f \otimes g)(x) = \int_{0}^{1} dy dz f(y)g(z) \delta(x-yz),
\ee
and we have implicitly summed over repeated parton indices.  The functions $\Gamma_{ij}$ are given by 
\be
\Gamma_{ij}(x) = \delta_{ij}\delta(1-x) -\frac{\alpha_s}{\pi} \frac{P^{(0)}_{ij}(x)}{\ep}.
\label{convkerns}
\ee
The DGLAP kernels $P^{(0)}_{ij}$ in the $\overline{\rm MS}$ scheme can be found in~\cite{Ellis:1991qj}; those 
required here are
\begin{eqnarray}
P^{(0)}_{qq}(x) &=& \frac{2}{3}\left\{\frac{3}{2}\delta(1-x)+\frac{2}{[1-x]_+}-(1+x)\right\}, \nonumber \\
P^{(0)}_{qg}(x) &=& \frac{1}{4}\left\{x^2+(1-x)^2\right\}.
\end{eqnarray}
To proceed, we substitute Eq.~(\ref{convkerns}) into Eq.~(\ref{newform}), equate this to Eq.~(\ref{baseform}), and 
solve for the renormalized cross sections ${\rm d} \hat{\sigma}_{ij}$.  We expand the renormalized cross sections 
in the strong coupling constant 
\be
{\rm d}\hat{\sigma}_{ij} = {\rm d}\hat{\sigma}^{(0)}_{ij} 
+ \left (\frac{\alpha_s}{\pi} \right) {\rm d}\hat{\sigma}^{(1)}_{ij},
\ee
and employ Eq.~(\ref{asexp}) to
obtain the following relations between the bare and renormalized cross sections at each order in $\alpha_s$:
\begin{eqnarray}
{\rm d}\hat{\sigma}^{(0)}_{q\bar{q}}(x_1,x_2) &=&  {\rm d}\sigma^{(0)}_{q\bar{q}}(x_1,x_2); \nonumber \\
{\rm d}\hat{\sigma}^{(1)}_{q\bar{q}}(x_1,x_2) &=& {\rm d}\sigma^{(1)}_{q\bar{q}}+\frac{1}{\ep}\int_{0}^{1}dy P^{(0)}_{qq}(y) 
  \left[{\rm d}\sigma^{(0)}_{q\bar{q}}(x_1,x_2 y)+{\rm d}\sigma^{(0)}_{q\bar{q}}(x_1 y,x_2)\right]; \nonumber \\
{\rm d}\hat{\sigma}^{(1)}_{qg}(x_1,x_2) &=& {\rm d}\sigma^{(1)}_{qg}+\frac{1}{\ep}\int_{0}^{1}dy P^{(0)}_{qg}(y) \,
  {\rm d}\sigma^{(0)}_{q\bar{q}}(x_1,x_2 y).
\label{colcount}
\end{eqnarray}
The collinear counterterms that must be added to the perturbatively computed cross sections are the integrals in Eq.~(\ref{colcount}).  
These are straightforward to compute as Laurent series in $\ep$, as we do for the real radiation cross sections in Eq.~(\ref{NLORepexp}).  
We note that the convolution variable $y$ maps onto the invariant mass $s_{345}$ in Eq.~(\ref{NLORinvs}).  This makes it simple 
to check analytically that the singularities in the real radiation cross section that occur as $\lambda_1 \to 0,1$ cancel.  
We also note that the ${\cal O}(\ep)$ 
term in the leading order cross section in Eq.~(\ref{LOepexp2}) 
contributes to the collinear counterterms at the finite level.  

\subsection{Virtual corrections}

\begin{figure}[htb]
     \begin{center}
      \begin{picture}(330,50)(0,0)
        \SetColor{Blue}
        \ArrowLine(0,50)(25,50)
        \ArrowLine(25,0)(0,0)
        \Line(25,0)(25,50)
        \Photon(25,0)(50,0){3}{4}
        \Photon(25,25)(50,25){3}{4}
        \Photon(25,50)(50,50){3}{4}
        \put(-10,-2){2}
        \put(-10,48){1}
        \put(58,-2){3}
        \put(58,48){5}
        \put(58,23){4}
        \GlueArc(25,37.5)(6,90,270){2}{3}

        \ArrowLine(100,50)(125,50)
        \ArrowLine(125,0)(100,0)
        \Line(125,0)(125,50)
        \Photon(125,50)(150,50){3}{4}
        \Photon(125,0)(150,0){3}{4}
        \Photon(125,25)(150,25){3}{4}
        \put(90,-2){2}
        \put(90,48){1}
        \put(158,-2){3}
        \put(158,48){5}
        \put(158,23){4}
        \GlueArc(125,25)(10,90,270){3}{4}

	\ArrowLine(200,50)(225,50)
        \ArrowLine(225,0)(200,0)
        \Line(225,0)(225,50)
        \Photon(225,50)(250,50){3}{4}
        \Photon(225,0)(250,0){3}{4}
        \Photon(225,25)(250,25){3}{4}
        \put(190,-2){2}
        \put(190,48){1}
	\put(258,-2){3}
        \put(258,48){5}
        \put(258,23){4}
        \Gluon(225,12.5)(205,50){3}{5}

        \ArrowLine(295,50)(325,50)
        \ArrowLine(325,0)(295,0)
        \Line(325,0)(325,50)
        \Photon(325,50)(350,50){3}{4}
        \Photon(325,0)(350,0){3}{4}
        \Photon(325,25)(350,25){3}{4}
        \put(285,-2){2}
        \put(285,48){1}
	\put(358,-2){3}
        \put(358,48){5}
        \put(358,23){4}
        \Gluon(305,0)(305,50){3}{6}

      \end{picture}
     \end{center}
\caption{\label{NLOVdiags} 
          Representative one-loop diagrams contributing to $q\bar{q} \to ZZZ$.}
\end{figure}
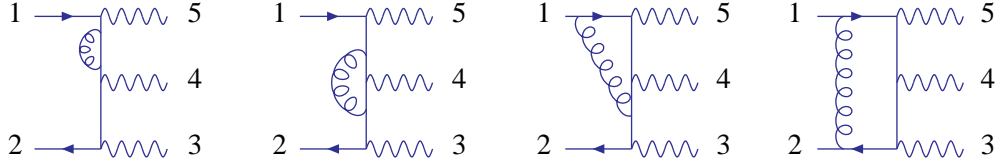

Finally, the virtual corrections to the partonic process $q\bar{q} \to VVV$ must be computed.  Forty-eight one-loop diagrams
contribute to this process; these must be interfered with 
the six
tree-level diagrams.  A representative sample of virtual diagrams 
is given in Fig.~(\ref{NLOVdiags}).  The expressions for the interferences are simple to obtain with standard techniques.  
We regulate all singularities using dimensional regularization, in which scale-less 
integrals are set to zero.  Consequently, there are no contributions from self-energy insertions on the external legs.  All 
singularities that remain when the one-loop diagrams are combined are infrared in origin.

The treatment of the virtual corrections to $2 \to 3$ processes is the main point of this paper, so we discuss this here
in detail.  We will use the pentagon appearing as the rightmost diagram in Fig.~(\ref{NLOVdiags}) to demonstrate our technique.  We begin by 
neglecting all numerator algebra that appears from the Feynman rules and Dirac traces when this diagram is interfered with 
the tree-level diagrams, and focus on the scalar topology.  The basic scalar integral 
for this topology is 
\be
{\cal I} = \int \frac{d^d k}{(2\pi)^d} \frac{1}{k^2}\frac{1}{(k+p_1)^2}\frac{1}{(k+p_3+p_4-p_2)^2}\frac{1}{(k+p_3-p_2)^2}\frac{1}{(k-p_2)^2}.
\ee
The $+i0$ prescription associated with each denominator has been suppressed for notational ease.  
After introducing a standard Feynman parameter representation, this integral becomes
\be
{\cal I} = -\frac{i \Gamma(3+\ep)}{(4\pi)^{d/2}} 
\int_{0}^{1} 
\prod \limits_{i=1}^{5} dx_i
%dx_1 dx_2 dx_3 dx_4 dx_5 
\,\delta\left(1-\sum_{n=1}^{5}x_i\right) [\Delta -i0]^{-3-\ep},
\label{pentint}
\ee
where
\begin{eqnarray}
\Delta &=& 2 x_2 (x_3+x_4) \,p_1 \cdot p_3 +2 x_2 x_3 
\,p_1 \cdot p_4 +2(x_1+x_2)(x_3+x_4) \,p_2 \cdot p_3 
+2 x_3 (x_1+x_2) \,p_2 \cdot p_4
\nonumber \\ && -2 x_2(x_3+x_4+x_5) \,p_1 \cdot p_2 
-2 x_3 (x_1+x_2+x_5) p_3 \cdot p_4 
-x_3 (x_1+x_2+x_4+x_5) \,M_Z^2 \nonumber \\ 
&& -(x_3+x_4)(x_1+x_2+x_5)\,M_Z^2.
\label{feynden}
\end{eqnarray}
%\begin{eqnarray}
%\Delta &=& s_{13} x_2 (x_3+x_4) 
%+ s_{14} x_2 x_3 + s_{23}(x_1+x_2)(x_3+x_4)
%+s_{24} x_3 (x_1+x_2) 
%  \nonumber \\ && - s_{12} x_2(x_3+x_4+x_5) 
%- s_{34} x_3 (x_1+x_2+x_5) 
%-x_3 (x_1+x_2+x_4+x_5) \,M_Z^2 \nonumber \\ 
%  && -(x_3+x_4)(x_1+x_2+x_5)\,M_Z^2,
%\label{feynden}
%\end{eqnarray}
%
We must discuss two features of this integral: the extraction of infrared singular terms and the treatment of internal thresholds.

We begin by considering the singular structure.  This integral exhibits infrared singularities as various combinations of $x_i$ approach 0 or 1.  
A convenient, easily automated prescription for extracting such singularities from loop integrals was presented in~\cite{Binoth:2000ps}.  We 
summarize here the salient features of this technique.
\begin{enumerate}
  \item To remove singularities that occur as $x_i \to 1$, we first split the integral into primary sectors.  There is a primary sector for each 
    Feynman parameter; for example, the primary sector associated with $x_1$ is obtained by making the following variable changes in 
    Eq.~(\ref{pentint}):
    \be
      x_j = x_1 x^{'}_{j},\,\, j=2,3,4,5.
    \ee
    The $\delta$-function is then used to remove the integration over $x_1$.  All singularities are mapped to $x_i=0$ by this split.
  \item After forming the primary sectors, all singular terms arise when one or multiple $x_i$ vanish.  We use sector decomposition
    to handle the cases where several $x_i$ go to zero.  We illustrate this technique on the following simple example:
    \be
       {\cal I}^{'} = \int_{0}^{1} dxdy \, \frac{1}{(x+y)^{2+\ep}}.
    \ee
    We split this integral into two regions, ${\cal I}^{'} = {\cal I}^{'}_{1}+ {\cal I}^{'}_{2}$.  The first region has $x>y$, while the 
    second has $y>x$.  In the first region 
we make the variable change $y = y^{'}x$, while in the second we 
use $x = x^{'}y$.  The integrals become
    \be
       {\cal I}^{'}_{1} = \int_{0}^{1} dxdy^{'} \, \frac{x^{-1-\ep}}{(1+y^{'})^{2+\ep}},\,\, 
       {\cal I}^{'}_{2} = \int_{0}^{1} dx^{'}dy \, \frac{y^{-1-\ep}}{(1+x^{'})^{2+\ep}};
    \ee
    all singularities now occur only when a single $x_i$ vanishes.
  \item The singularities arising from $x_i^{-1-\ep}$ can be extracted using the plus distribution expansion in Eq.~(\ref{plusexp}).  This yields 
    a Laurent series in $\ep$ whose coefficients can be integrated either analytically or numerically.  In some cases it is convenient 
    to modify the Feynman parameterization in Eq.~(\ref{feynden}) to reduce the number of sector decompositions required.
\end{enumerate}

The Feynman denominator $\Delta$ in Eq.~(\ref{feynden}) can vanish in the interior of the $x_i$ integration region, as it is clear from the 
presence of terms with both plus and minus signs.  This occurs when the internal loop particles go on-shell, and signals the onset of an 
imaginary part in the integral.  The $-i0$ prescription regulates these internal thresholds.  However, this prescription is not suitable 
for a numerical treatment of the integral.  A method that allows internal thresholds to be handled completely numerically was 
developed in~\cite{Soper,Soper2}.  The idea is to deform the contour for the Feynman parameter integrations away from the real axis in the direction 
indicated by the $-i0$ term.  If the contour is sufficiently far from where $\Delta$ vanishes, then the integration can be performed 
numerically in the complex plane.  

We discuss the application of this technique to the pentagon integral 
in Eq.~(\ref{pentint}).
After splitting the integral into primary sectors and 
sector decomposing the integrand, the 
integrand denominator in each sector 
is given by a product of Feynman parameters factored out during 
the sector decomposition procedure, and a function $\tilde \Delta$ that 
depends upon kinematic variables. This function may vanish in the 
interior of the integration region 
and, for this reason, its properties determine the desired contour deformation. The function $\tilde 
\Delta$ 
takes on the generic form
\be
\tilde \Delta = Z +\sum_{i} Y_i x_i +\sum_{i,j} \frac{1}{2}X_{ij}x_i x_j+\sum_{i,j,k} \frac{1}{3}W_{ijk}x_i x_j x_k +\ldots \, .
\ee
The tensors $X$, $Y$, and $W$ consist of kinematic invariants and are independent of the Feynman parameters.  
The ellipsis denotes terms of quartic and higher order in the Feynman parameters $x_i$.  Terms beyond quadratic order appear only for 5- or 
higher-point integrals, and only after the sector decomposition is performed.  We find that it is necessary to perform the sector 
decomposition before deforming the integration contour; reversing the order 
can lead to thresholds regulated only by $-i0$.  The idea is to now set 
$x_i = y_i -i\tau_i$, and choose $\tau_i$ such that  internal thresholds are avoided and the integration over $y_i$ can be done numerically.  
A convenient choice for $\tau_i$, similar to that presented in~\cite{Soper2}, is
\be
\tau_i = \lambda y_i (1-y_i) \left[Y_i + \sum_{j} X_{ij}y_j + \sum_{j,k} W_{ijk} y_j y_k +\ldots \right].
\label{deform}
\ee
The endpoints of the contour remain fixed with this choice.  The parameter
$\lambda$ controls the overall size of the deformation.  Because 
$\tilde \Delta$
contains cubic and higher polynomials in $x_i$, the deformation in Eq.~(\ref{deform}) does not ensure a sign-definite imaginary part.  However, 
as $\lambda \to 0$ the deformation is in the direction required by the $-i0$ prescription; this allows us to begin with a small choice 
of $\lambda$, and check that no pole in the complex plane is crossed as we increase 
$\lambda$.  For numerical purposes, it is 
convenient to choose $\lambda$ large as compared to the kinematic invariants in $X$, $Y$, and $W$.  It is simple to obtain the kinematic matrices 
using computer algebra techniques, indicating that the 
contour deformation procedure can be completely automated.

Computing the virtual corrections is simple once the singularity and threshold structures of the base scalar integrals are regulated.  The 
numerator algebra which arises from computing a complete diagram has the generic form ${\cal N}(k^2,k \cdot p_i)$ in terms of the loop momentum.  
After analytically integrating over $k$ using standard techniques, the numerator becomes a polynomial in the Feynman parameters;
it can be treated 
numerically.  No reduction of tensor structures is needed.  We found it convenient to put each diagram over a common Feynman denominator; large 
cancellations between tensor and scalar integrals occur if the tensor integrals are sector decomposed separately.  Each interference between 
one-loop
and tree diagrams becomes a Laurent series in $\ep$, with coefficients that can be integrated numerically.  Judicious grouping 
of terms allows the expression size for each diagram to be kept relatively compact.

In summary, our procedure for computing the virtual corrections is as follows.
\begin{enumerate}
  \item Compute the interference between a 
one-loop
diagram and the tree diagrams using standard techniques.
  \item Identify the base scalar integral for each diagram.  Introduce a Feynman parameterization for this integral, and combine
    all tensor structures into a numerator over the denominator of the scalar integral.  Perform the integration over loop momentum 
    analytically; the tensor terms become polynomials in Feynman parameters.
%simple polynomials in the $x_i$.
  \item Split the base integral into primary sectors, and sector decompose each primary sector until all singularities are extracted.
    The resulting expression will be a Laurent series in $\ep$ with 
integrable
coefficients.
  \item Deform the contour in each sector by making the variable change in Eq.~(\ref{deform}).  This regulates all internal thresholds and allows
for a numerical evaluation of the integral.
\end{enumerate}
%
%At this point, the virtual corrections can be combined with the real 
%radiation and collinear counterterms, and the cancellation of 
%all $1/\ep$ poles can be established.  We have checked that the 
%partonic channels $q\bar{q} \to ZZZ+X$ and $qg \to ZZZ+X$ are 
%indeed finite when all components are combined.

\section{Results}
\label{sect4}

In this Section we describe the results of our computation of the NLO 
QCD corrections to the tri-boson production process $pp \to ZZZ$ at the LHC.
We employ the following numerical values 
for the Fermi constant and the weak boson masses:
\be
G_F = 1.166 \times 10^{-5} \,{\rm GeV}^{-2}, \,\,M_W = 80.451 \,{\rm GeV}, \,\, M_Z = 91.1875 \,{\rm GeV}.
\ee
We use the MRST~\cite{mrst} parton distribution functions 
at either LO or NLO, as appropriate.  The values of the strong coupling 
constant $\alpha_s(M_Z)$ appropriate to use with the MRST parton distribution 
functions are also obtained from Ref.~\cite{mrst}.

We compute the real emission corrections using the procedure described in 
Section~\ref{sect3}. To compute the virtual  
corrections, we generate a set of ten thousand
random kinematic events distributed according to the Born-level matrix element and the leading 
order parton distribution functions.
Each event is described by seven variables that provide 
a complete description of the $q \bar q \to ZZZ$ kinematics.  We compute the NLO virtual correction for each 
event and then reweight the event using the computed correction 
and the ratio of NLO to LO parton distribution functions.

For the numerical computation of the NLO corrections, we employ the adaptive 
Monte Carlo integration program VEGAS as implemented in the CUBA library~\cite{cuba}.
The numerical stability of the computation is exceptional; all diagrams 
including the pentagons exhibit a very fast rate of convergence. 
The computation of the NLO QCD corrections for ten thousand 
kinematic points required a few days of running on a cluster  
of several dozen processors. 

As an example of our code output we present below in Table~\ref{table1} a listing of the finite contributions 
coming from the one-loop corrections
for several sample events.  The kinematics are defined by $x_1,x_2$, the values of Bjorken-$x$ for each proton, and the 
kinematic invariants $s_{ij} = (p_i-p_j)^2$.  The initial partonic momenta are denoted by $p_1, p_2$ , while $p_3,p_4$ indicate
the final-state $Z$-momenta.  The invariant masses have been scaled by $1/(x_1x_2s)$ so that their magnitude is between zero and one.  
All other invariant masses can be obtained via momentum conservation.  The quantity $z=M_Z/\sqrt{x_1x_2s}$ is included for completeness.  
Each event has unit weight at leading order.  The shift in the weight coming from the NLO virtual corrections is obtained via
\begin{equation}
w_{{\rm NLO}} = \frac{{\rm PDF(NLO)}}{{\rm PDF(LO)}}\left(1 + \frac{\alpha_s}{\pi}V\right).
\end{equation}
$V$ is the result from numerically evaluating the loop corrections, and is given in the table for each event.  These must be 
combined with the real corrections and collinear counterterms 
to produce the final result.  We generate these remaining contributions in our program as separate events.  Both the parton distribution 
functions and $\alpha_s$ can be evaluated at the desired scales.  We note that the events have been generated using the factorization 
scale choice $\mu_F=3 M_Z$ at leading order.

\begin{tiny}
\begin{table}[htbp]
\vspace{0.1cm}
\begin{center}
\begin{tabular}{|c|c|}
\hline\hline
$\left\{x_1,x_2,z,s_{13},s_{23},s_{14},s_{24}\right\}$ & $V$ \\ \hline\hline
$\left\{0.182,0.030,0.088,-0.640,-0.143,-0.132,-0.802\right\}$ & $7.59(2)$ \\ \hline
$\left\{0.138,0.006,0.226,-0.293,-0.240,-0.487,-0.095\right\}$ & $9.76(2)$ \\ \hline
$\left\{0.024,0.193,0.096,-0.035,-0.569,-0.692,-0.065\right\}$ & $9.57(7)$ \\ \hline
$\left\{0.032,0.052,0.160,-0.423,-0.281,-0.226,-0.152\right\}$ & $7.93(2)$ \\ \hline
$\left\{0.014,0.074,0.202,-0.387,-0.060,-0.040,-0.750\right\}$ & $11.77(4)$ \\ \hline\hline
\hline
\end{tabular}
\caption{\label{table1} Finite contributions coming from the one-loop corrections for several sample events.  The first column indicates 
the kinematics of the events, while the second column gives the NLO virtual correction with its associated Monte Carlo 
integration error.  The notation is as defined in the text.}
\vspace{-0.1cm}
\end{center}
\end{table}
\end{tiny}

We have applied a number of checks to our calculation.  
\begin{enumerate}
\item We have compared 
the leading order cross-section obtained with our code with 
the result of a similar computation using the program 
MadEvent~\cite{Maltoni:2002qb} and have found complete agreement.

\item As we mentioned earlier, the NLO virtual corrections are divergent, and 
physical results are only obtained once real emission contributions are added. The divergent part of the NLO virtual correction to 
$ q(p_1) + \bar q(p_2) \to ZZZ$ is related 
to the leading order cross-section  by the following equation~\cite{catani}:
\be
\sigma^{\rm NLO, virt}|_{\rm div} = 
-C_F \frac{\alpha_s}{\pi} \frac{\Gamma(1+\ep)}{(4\pi)^{-\ep}} (s_{12})^{-e}
\left ( \frac{1}{\ep^2} +\frac{3}{2\ep} \right ) \sigma^{\rm LO},
\label{eqcat}
\ee
where $s_{12} = 2p_1\cdot p_2$ and $C_F = 4/3$ is the QCD color factor.  
We have checked that our numerical
computation of  $\sigma^{\rm NLO, virt}$ gives the divergent part in 
full agreement with Eq.~(\ref{eqcat}).

\item We have checked that all divergences 
cancel at the differential level once the real emission processes, 
the collinear counterterms and the virtual corrections are combined.

\item An important check of the result is provided by its independence 
of $\lambda$, the size of the contour deformation. However, we stress that 
the efficiency of the numerical integration  depends 
strongly on $\lambda$.  For small values of $\lambda$ one does not 
move sufficiently far  from the pole on the real axis, while for 
large values of $\lambda$ one deforms too much and there are large 
cancellations between different segments of the integration path in the complex plane.

\item Finally, we have implemented 
all parts of the computation in at least 
two independent computer codes that agree for all observables studied.  We have computed the real emission 
processes using both the approach described in the text and the traditional phase-space slicing method, and 
have found complete agreement.

\end{enumerate}

\noindent
\begin{figure}[htbp]
\vspace{0.0cm}
\centerline{
\includegraphics[height=7.5cm,width=6.0cm,angle=90]{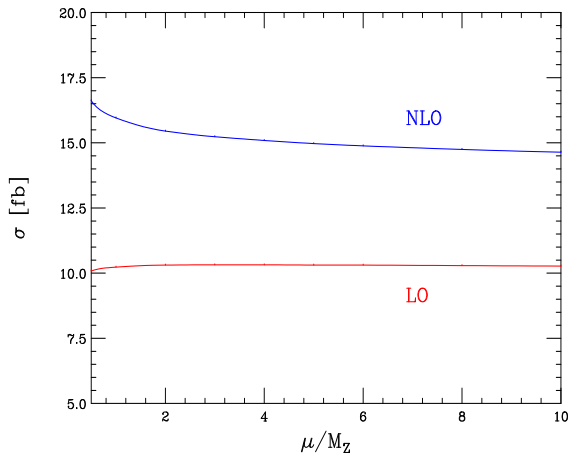}
}
\caption{The scale-dependence of the leading order and next-to-leading 
order cross-sections $\sigma(pp \to ZZZ)$. We have 
set the factorization and the renormalization scales equal to a common
value $\mu$.
}
\label{rscale}
%\vspace{-0.3cm}
\end{figure}

\noindent
\begin{figure}[htbp]
\vspace{0.0cm}
\centerline{
\includegraphics[height=7.5cm,width=6.0cm,angle=90]{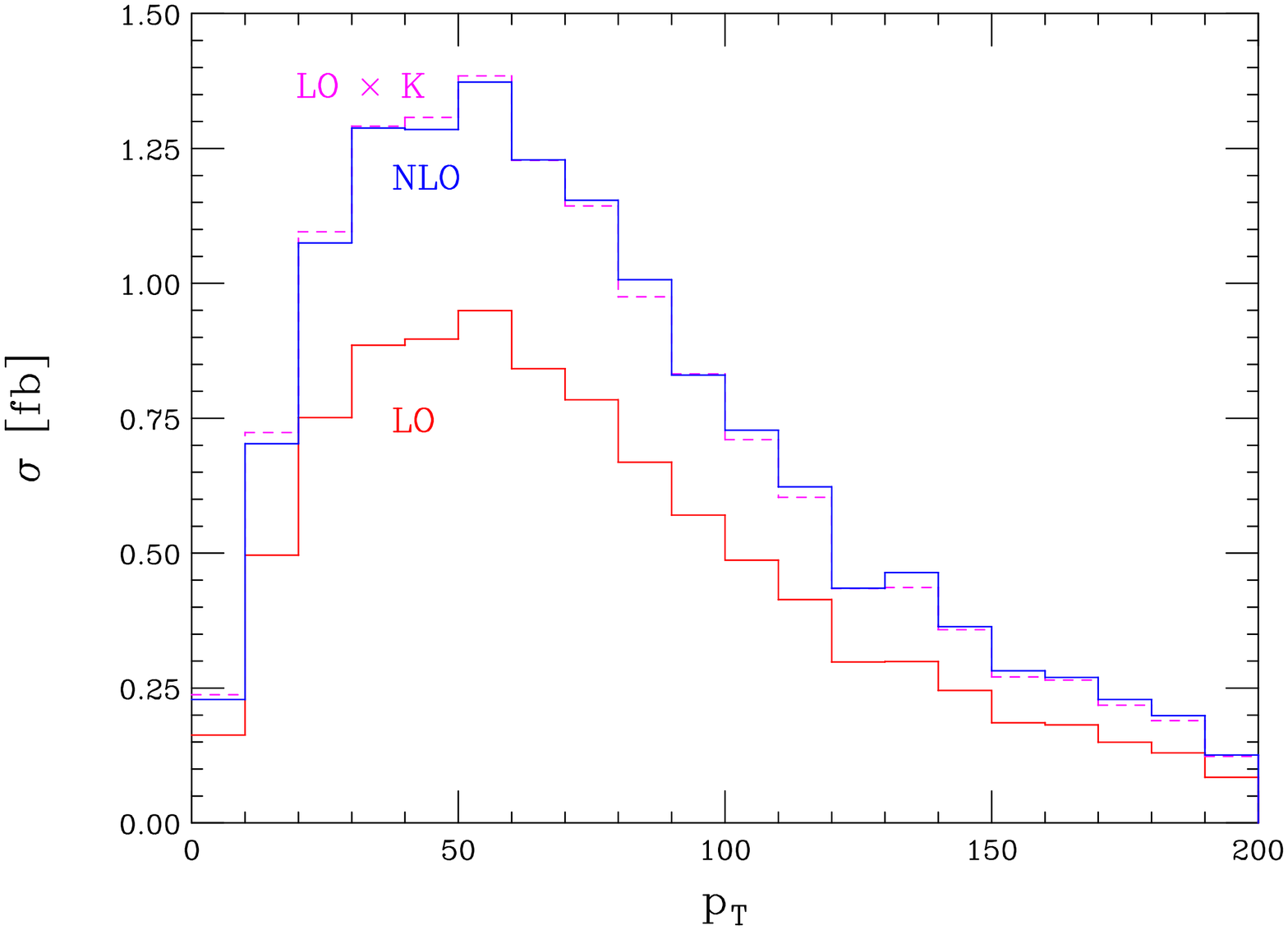}
\includegraphics[height=7.5cm,width=6.0cm,angle=90]{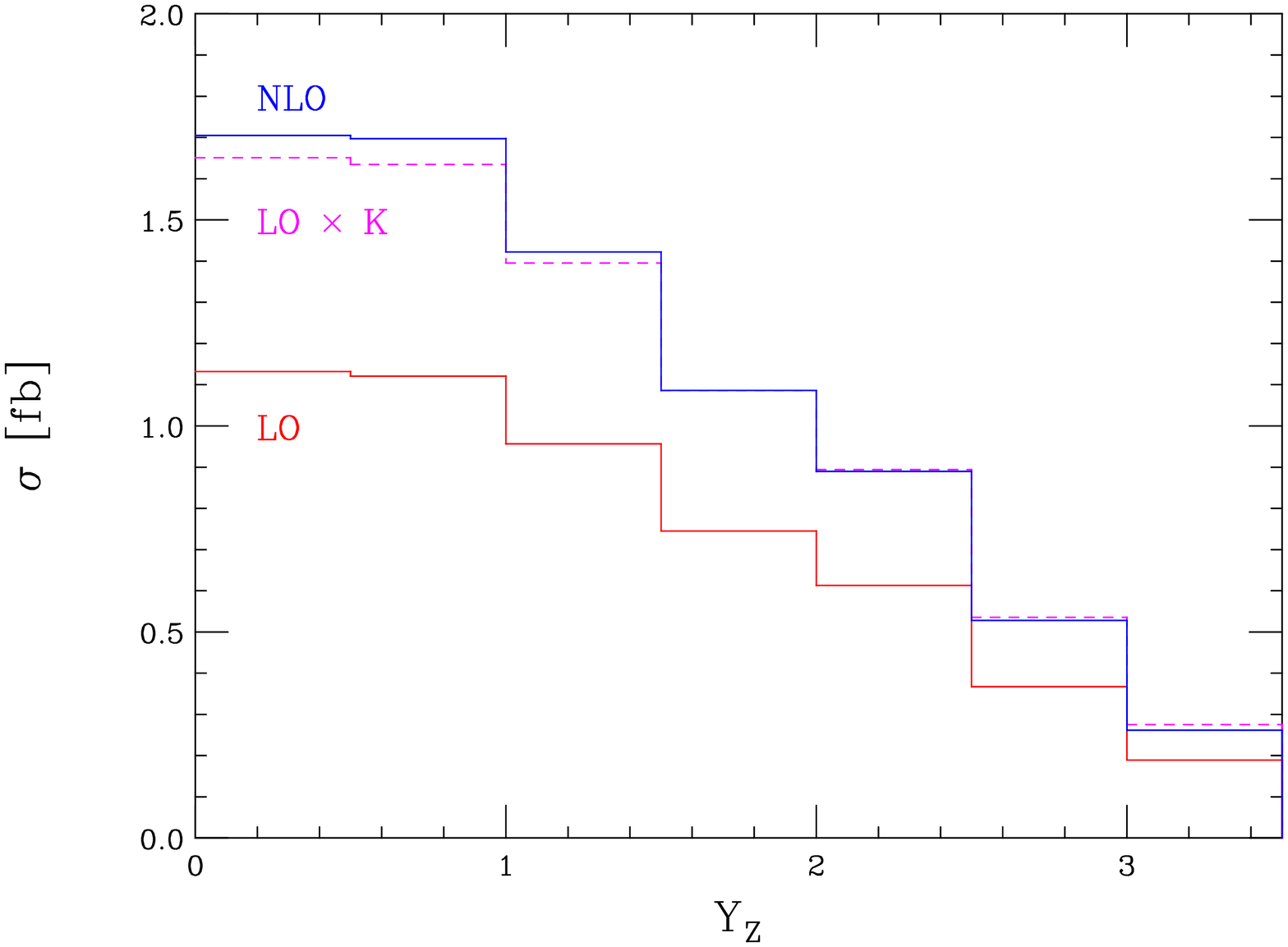}
}
\caption{The transverse momentum and rapidity distributions of the 
$Z$ bosons at LO and NLO in $\alpha_s$, normalized by a factor $1/3$. 
The results obtained by re-scaling the LO distribution by a constant $K$-factor are also shown.
The value of the factorization and the renormalization scales are 
set equal to $3 M_Z$.
}
\label{rdistr}
%\vspace{-0.3cm}
\end{figure}

We now describe the results of our computation of the NLO QCD corrections 
to $pp \to ZZZ$.  In Fig.~\ref{rscale} we show the dependence of the total cross-section
computed through leading and next-to-leading order on the renormalization and 
factorization scales.  We have equated these to a common scale $\mu=\mu_R=\mu_F$.  There 
are two important features of this result to note.  The first is that the corrections are 
large, approximately 50\% over a wide range of $\mu$.  This results from a large 
increase in the $q\bar{q}$ luminosity function when going from LO to NLO, and 
large virtual corrections.  For example, for $\mu=3 M_Z$, the LO cross section evaluated 
with LO parton distribution functions is 10.3 fb, while the LO cross section evaluated 
with NLO distribution functions is 11.4 fb.  The full NLO result is 15.2 fb, with the 
additional increase coming entirely from the virtual corrections.  The 
effect of real parton emission in the $qg$ and $\bar q g$ channels is 1\% or less for 
all $\mu$ considered.  We note that similarly large corrections for the process $pp \to ZZ$ at the LHC 
were observed in~\cite{zz}.

The second important feature is the tiny scale dependence of the LO result, which 
drastically underestimates the NLO correction.  The LO result varies by only a few percent over the 
entire range of $\mu$ considered.  While such behavior is uncommon, it is by no means 
unique to this process; a very similar situation occurs for $Z$ production 
at the Tevatron~\cite{zrap}.  

In Fig.~\ref{rdistr} we present the transverse 
momentum and rapidity distributions of the $Z$ bosons.  We include all three bosons and divide by a factor of 3 to
normalize the result.  We compare these distributions to the approximation of reweighting the LO results by a constant 
$K$-factor, where $K$ is the ratio of NLO to LO inclusive cross sections.  For the 
distributions studied, the NLO QCD corrections do not depend 
significantly on the kinematics of the produced particles.  Rescaling the leading order kinematic 
distributions by a constant $K$-factor gives a description of the NLO result accurate to a few percent.  
We expect that this is true in all kinematic regions for which phase space is available at 
leading order.

\section{Conclusions}
\label{sect5}

In this paper we present a novel numerical  
method for perturbative computations 
in QCD to next-to-leading order accuracy. 
We combine sector decomposition~\cite{Binoth:2000ps},
which allows an  automatic extraction of soft and collinear 
singularities from virtual diagrams, with 
contour deformation of the Feynman parametric integrals~\cite{Soper,Soper2}, which permits 
numerical evaluation of loop corrections with internal thresholds.
Doing so, we obtain a tool that enables an efficient and flexible numerical 
evaluation of Feynman diagrams with an arbitrary singularity structure.
It appears to us that this combination of sector decomposition and contour deformation provides a framework 
in which numerical
calculations of NLO virtual corrections can be fully automated.

To test this idea, we compute the next-to-leading order QCD corrections to the 
production of three $Z$ bosons in proton-proton collisions; this is 
one of the processes that appears on the so-called ``NLO wishlist''
\cite{NLOwish}. We observe that the method possesses excellent efficiency and numerical stability; 
for all phase-space points considered, we were able 
to compute the NLO virtual corrections with sub-percent precision.

The NLO QCD corrections to $pp \to ZZZ$ are large, approximately 50\% for all scale 
choices considered.  The leading order scale 
dependence drastically underestimates the size of these corrections.  For phase space regions 
accessible at leading order, the NLO corrections are independent of the kinematics of the final state particles.

\medskip

{\bf Acknowledgments}: 
K.M. is supported in part by the DOE grant DE-FG03-94ER-40833, Outstanding 
Junior Investigator Grant and by the Alfred P.~Sloan Foundation. A.L. is supported with funds 
provided by the Alfred P. Sloan Foundation.
F.P. is supported in part 
by the DOE grant DE-FG02-95ER40896, by the University of Wisconsin Research Committee 
with funds provided by the Wisconsin Alumni Foundation, and 
by the Alfred P.~Sloan Foundation. 

%\thispagestyle{empty}
%\end{titlepage}

\end{document}